\documentclass[showpacs,twocolumn,pra,superscriptaddress,notitlepage]{revtex4-1}
\usepackage{qcircuit}
\usepackage[dvips]{graphicx}
\usepackage{amsmath,amssymb,amsthm,mathrsfs,amsfonts,dsfont}
\usepackage{subfigure, epsfig}
\usepackage{braket}
\usepackage{bm}
\usepackage{enumerate}
\usepackage{color}

\newcommand{\tr}{\mathrm{Tr}}

\newcommand{\brakett}[2]{\langle#1 \vert #2 \rangle}
\usepackage{color}
\newcommand{\PKU}{Center on Frontiers of Computing Studies, Peking University, Beijing 100871, China}

\newcommand{\PKUCS}{School of Computer Science, Peking University, Beijing 100871, China}

\begin{document}

\title{A resource-efficient quantum-classical hybrid algorithm for energy gap evaluation}

\author{Yongdan Yang}
\affiliation{Graduate School of China Academy of Engineering Physics, Beijing 100193, China}

\author{Ying Li}
\affiliation{Graduate School of China Academy of Engineering Physics, Beijing 100193, China}

\author{Xiaosi Xu}
\email{xsxu@gscaep.ac.cn}
\affiliation{Graduate School of China Academy of Engineering Physics, Beijing 100193, China}

\author{Xiao Yuan
}
\email{xiaoyuan@pku.edu.cn}

\affiliation{\PKU}
\affiliation{\PKUCS}

\begin{abstract}
{Estimating the eigenvalue or energy gap of a Hamiltonian $H$ is vital for studying quantum many-body systems. Particularly, many of the problems in quantum chemistry, condensed matter physics and nuclear physics investigate the energy gap between two eigenstates. Hence, how to efficiently solve the energy gap becomes an important motive for researching new quantum algorithms. In this work, we propose a hybrid non-variational quantum algorithm that uses the Monte Carlo method and real-time Hamiltonian simulation to evaluate the energy gap of a general quantum many-body system. Compared to conventional approaches, our algorithm does not require controlled real-time evolution, thus making its implementation much more experimental friendly. Since our algorithm is non-variational, it is also free from the “barren plateaus” problem. To verify the efficiency of our algorithm, we conduct numerical simulations for the Heisenberg model and molecule systems on a classical emulator.}
\end{abstract}
\maketitle

\section{Introduction}

Solving the many-body Schrodinger’s equation remains a challenge across various fields. In the past years, efforts have been made to develop methods for a solution to the many-body systems. However, using classical methods to handle this task becomes increasingly difficult as the size of the system increases. Quantum computing arises as an alternative to solve such problems with the potential to surpass its classical counterpart. The quantum phase estimation (QPE) algorithm~\cite{nielsen2002quantum,kitaev1995quantum} is one of the most influential quantum algorithms that can find the energy eigenvalues of a given Hamiltonian. However, implementing the algorithm requires a deep circuit and thus makes it infeasible for near-term quantum computers~\cite{preskill2018quantum}. In recent years, many new methods suitable for the noisy intermediate-scale quantum devices have been developed, most target finding the ground state of a given Hamiltonian, such as quantum subspace diagonalization~\cite{colless2018computation,epperly2022theory,cohn2021quantum,cortes2022quantum}, imaginary time evolution~\cite{huo2021shallow,motta2020determining,lin2021real,mcardle2019variational}, variational algorithms~\cite{peruzzo2014variational,kandala2017hardware,wang2019accelerated,wecker2015progress}, hybrid quantum Monte Carlo methods~\cite{huggins2022unbiasing,xu2022quantum,zhang2022quantum}, etc. These methods come with unique advantages yet also limitations. For example, some algorithms mentioned above are based on variational principles for finding the ground-state energy. Those algorithms suffer from the “barren-plateau” problem~\cite{mcclean2018barren} and their efficiency largely depends on the ansatz. There is a range of other algorithms~\cite{somma2019quantum,o2019quantum,clinton2021phase} that have a more ambitious goal than the ground-state energy estimation and aim to estimate the distribution of all eigenvalues started from a given initial state $\ket{\psi_0}$. However, implementation of these algorithms at a meaningful scale still requires a fault-tolerant quantum computer.

On the other hand, finding the energy gap of a physical system, especially in the low-lying subspace, is important for fields including quantum chemistry~\cite{sugisaki2021bayesian,hunt2018quantum,shee2019singlet,cao2019quantum}, condensed matter physics~\cite{shi2016energy,capelle2007energy}, and nuclear physics~\cite{li2009microscopic,zhu2003composite,hoffman2008determination}. To find the difference between different energy levels, one approach is  to evaluate the energy of the corresponding eigenstates and then deduct the two. Doing so requires more quantum resources, and the results may not be accurate with the presence of noise. Hence, it is important to explore more resource efficient quantum algorithm that can directly evaluate the energy gap.

This paper introduces a new hybrid non-variational algorithm that can efficiently find energy gaps in the low-lying subspace. While there exists prior research \cite{sugisaki2021bayesian,matsuzaki2021direct,sugisaki2021quantum} that focuses on estimating energy gaps, our approach follows a distinct paradigm. In our method, we use quantum computing to simulate real-time evolution, and the classical Monte Carlo method is used to evaluate a Fourier integral. Hence, our method is free from the sign problem. Our method could be regarded as an extension of the recently proposed algorithms~\cite{Ge19,lu2021algorithms,CVLCU,Rodeo21,lin2021heisenberglimited,zeng2021universal}, which applies Fourier transform to more efficiently estimate Hamiltonian eigenstates and eigenenergies. Our work introduces a more general framework that can both evaluate energy eigenstates and gaps. To find the energy gaps, we show that our method uses even shallower circuits without requiring to implement the controlled time-evolution. 
It is thus is more practical for near-term and early fault-tolerant devices. One special case of our method can also be used to find energy eigenstates, which reduced to the case discussed in Ref.~\cite{zeng2021universal}. 

This paper is organized as follows. In Sec. \uppercase\expandafter{\romannumeral2} and Sec. \uppercase\expandafter{\romannumeral3}, we introduce our method to evaluate the energy gaps and energy eigenvalues for a given Hamiltonian $H$ respectively. In Sec. \uppercase\expandafter{\romannumeral4}, we discuss the case with degenerate energy levels. We present the numerical simulation results in Sec. \uppercase\expandafter{\romannumeral5}, and 
perform error analysis in Sec. \uppercase\expandafter{\romannumeral6}. Sec. \uppercase\expandafter{\romannumeral7} is the conclusion of this paper.

\section{Energy gap evaluation}
To better illustrate our idea, we introduce our method with the assessment of the energy gaps for a given Hamiltonian $H$. The time evolution operator in the Schr\"{o}dinger picture writes $e^{-iHt}$, which can be decomposed into a diagonal matrix  $e^{-iHt}=\sum_ie^{-iE_it}\ket{i}\bra{i}$ with energy eigensvalues $E_i$ and eigenstates $\{\ket{i}\}$. To extract the information of the energy gaps from this operator, we consider $C(E)$ as the Fourier transform of function $f(t)$
\begin{align}
C(E)=\int_{-\infty}^\infty f(t)e^{iEt}dt,
\label{eq:c(x)}
\end{align}
with $f(t)$ a function contains $e^{-iHt}$ as
\begin{align}
f(t)=p(t)\tr[O\ket{\psi(t)}\bra{\psi(t)}],
\end{align}
where $\ket{\psi(t)}=e^{-iHt}\ket{\psi_0}~(\hbar=1)$ and $\ket{\psi_0}$ is the initial state. Here, $p(t)$ is a cooling function adopted from Ref.~\cite{zeng2021universal} to 
guarantee that $C(E)$ can be efficiently sampled within a finite sampling range, which we will discuss later. As defined, $p(\tau)$ monotonically increases as $\tau$ increases and satisfies $\rm{lim}_{E\to\infty}p(\tau E)/p(\tau E^{'})=0$ for any $E>E^{'}>0$. 

The function $C(E)$ could be rewritten as follows
\begin{equation}
\begin{aligned}
&C(E)=\int_{-\infty}^\infty p(t) \tr[O\ket{\psi(t)}\bra{\psi(t)}]e^{iEt}dt\\
&\hspace{-2.5mm}=\tr\left[O\sum_{i,j=0}\ket{i}\brakett{i}{\psi_0}\brakett{\psi_0}{j}\bra{j}\int_{-\infty}^\infty e^{-i(E_i-E_j-E)t}p(t)dt\right]
\label{eq:c(e)}
\end{aligned}
\end{equation}
Note that when $p(t)$ is a Lorentz function $p(t)=\frac{\beta}{\pi}\frac{1}{\beta^2+t^2}$, its Fourier transform $\mathcal{P}(\omega)=\int_{-\infty}^\infty p(t)e^{i\omega t}dt = e^{-\beta|\omega|}$ has exactly the same form of imaginary evolution for $\omega>0$. In general, $p(t)$ could be any function that satisfies the convergent condition described above. Since it is shown that the Gaussian function has the best performance among several other example functions, in this work, we also assume that $p(t)$ is 
the Gaussian function as an example in the further analysis, i.e., 
\begin{align}
p(t)=e^{-a^2t^2},
\end{align}
where $0<a<1$ is a tuning parameter.

Therefore, Eq.~(\ref{eq:c(e)}) becomes
\begin{align}
C(E)=\tr\left[O\sum_{i,j=0}\ket{i}\brakett{i}{\psi_0}\brakett{\psi_0}{j}\bra{j}\frac{\sqrt{\pi}}{a}e^{-\frac{[E-(E_i-E_j)]^2}{4a^2}}\right].
\label{eq:c(e)final}
\end{align}
By gradually increasing $E$ from 0, $C(E)$ reaches the maximum when $E=E_i-E_j$. For convenience, in the following discussions we assume $E_i>E_j$.

One could tell from the expression of $C(E)$ that two conditions should hold to make it work. (1) The overlaps between the initial state and the eigenstates $\ket{i}$ and $\ket{j}$ must not be small. (2) The observable $O$ should be carefully chosen such as $\langle i|O|j\rangle$ is not small. In most of the applications where people are interested in the energy gap between two eigenstates~\cite{sugisaki2021bayesian,hunt2018quantum,zhu2003composite,hoffman2008determination}, the energy of one eigenstate is roughly known, thus a natural choice for the initial state $\ket{\psi_0}$ is the estimated $i$-level eigenstate. In such case, the overlaps satisfy $\brakett{i}{\psi_0}\gg\brakett{j}{\psi_0}$ for all $i\neq j$. By doing so, the energy gaps between level $i$ and all other levels stand out in the full energy eigenvalues, making it easier to identify the gaps between $i$ and other levels. Following the same logic, we can make the observable $O=\ket{\psi_0}\bra{\psi_0}$, therefore $\langle i|O|j\rangle$ is transformed into the overlaps between the eigenstates and initial state.

We can then summarize our method as follows
\begin{enumerate}
    \item Choose the time $t$ randomly according to $p(t)$.
    \item Evolve the quantum state with Hamiltonian $H$ and time $t$, measure $O$ to obtain $\tr[O\ket{\psi(t)}\bra{\psi(t)}]$, and multiply it by $e^{-Et}$ for parameter $E$.
    \item Repeat 1 \& 2 and compute the average result to obtain $C(E)$.
\end{enumerate}
Note that the state $\ket{\psi(t)}=e^{-iHt}\ket{\psi_0}$ can be efficiently prepared with Trotterization~\cite{huo2021shallow} using a quantum computer. While the integral for $t$ is from $[-\infty,\infty]$, we could consider a cutoff to be within a finite range $[-T,T]$ with a sacrifice of an error $\epsilon\sim \mathcal O(e^{-(aT)^2})$. 
The detailed error analysis can be found later.

\vspace{0.4cm}

So far we have considered the case to find the energy gaps between different energy levels of the same physical system described by a single Hamiltonian. In many cases, especially condensed matter physics~\cite{shi2016energy}, people are interested to find out energy differences between two eigenstates of the physical system under different conditions, i.e., with two different Hamiltonians $H^{(1)}$ and $H^{(2)}$. Our method also fits this scenario. We need to rewrite the function $f(t)$ into
\small
\begin{equation}
\begin{aligned}
&f(t)=p(t)\tr\left[Oe^{-iH^{(1)}t}\ket{\psi_0}\bra{\psi_0}e^{iH^{(2)}t}\right]\\
&\hspace{-2mm}=p(t)\tr\left[O\sum_{m,n}e^{-i(E^{(1)}_m-E^{(2)}_n)t}\ket{m^{(1)}}\brakett{m^{(1)}}{\psi_0}\brakett{\psi_0}{n^{(2)}}\bra{n^{(2)}}\right].
\label{eq:two_H}
\end{aligned}
\end{equation}
\normalsize
Here $\{\ket{m^{(1)}}\}$ and $\{\ket{n^{(2)}}\}$ are eigenstates of the Hamiltonian $H^{(1)}$ and $H^{(2)}$ respectively.

As we are interested in the energy difference of the two eigenstates, they should have a non-zero overlap to $\ket{\psi_0}$ to make the method work. The initial state $\ket{\psi_0}$ could be an estimation of one of the eigenstates, and $O$ can be chosen as $\ket{\psi_0}\bra{\psi_0}$.
By doing so, we need the quantum computer to evaluate the transition function $\tr[Oe^{-iH^{(1)}t}\ket{\psi_0}\bra{\psi_0}e^{iH^{(2)}}t]$, which can be realized with the Hadamard test circuit (needs controlled-evolution) or with a simple circuit with an ancilla if $\tr[O\ket{\psi_0}\bra{\psi_0}]$, $\tr[O\ket{\psi_0}\bra{\psi(t)}]$ and $\tr[O\ket{\psi(t)}\bra{\psi_0}]$ are known, as proposed in Ref.~{\cite{huo2021shallow}}.

\section{Energy evaluation}

To find the energy gaps, we write $C(E)$ in Eq.~(\ref{eq:c(x)}) as a function with a single variable dependent on the energy difference of two levels. We could also write the function 
\begin{align}
C(x,x^{'})=\iint_{-\infty}^\infty f(t,t^{'})e^{ixt}e^{ixt^{'}}dtdt^{'},
\label{eq:c(x,x)}
\end{align}
where $f(t,t^{'})=p(t)p'(t)\tr[O\ket{\psi(t)}\bra{\psi(t^{'})}]$ such as one can also evaluate the energies instead of the energy gaps by tuning the parameters $E, E^{'}$:
\small
\begin{equation}
\begin{aligned}
\hspace{-2mm}&C(E,E^{'})\\
&=\tr\left[O\sum_{i,j=0}\ket{i}\brakett{i}{\psi_0}\brakett{\psi_0}{j}\bra{j}\frac{\pi}{a^2}e^{-\frac{(E-E_i)^2}{4a^2}}e^{-\frac{(E^{'}-E_j)^2}{4a^2}}\right].
\label{eq:c(x,x)_2}
\end{aligned}
\end{equation}
\normalsize
Here we have chosen $p(t)=p'(t)=e^{-a^2t^2}$. But we note that $p(t)$ and $p'(t)$ does not need to be the same function, instead they could be different functions for better differentiating the two energies. Besides, with $E'=0$, Eq.~(\ref{eq:c(x,x)}) can be simplified as 
\begin{align}
C(E,0)=\int_{-\infty}^\infty f(t,0)e^{iEt}dt,
\label{eq:c(x,0)}
\end{align}
where $f(t,0)=p(t)\tr[O\ket{\psi(t)}\bra{\psi_0}]$, such as we have 
\begin{align}
C(E,0)=\tr\left[O\sum_{i=0}\ket{i}\brakett{i}{\psi_0}\bra{\psi_0}\frac{\sqrt{\pi}}{a}e^{-\frac{(E-E_i)^2}{4a^2}}\right].
\label{eq:energy_direct}
\end{align}
This enables us to obtain the whole energy eigenvalues of the Hamiltonian. In this case, we can make $O=I$ such that $\tr[\sum_{i=0}\ket{i}\brakett{i}{\psi_0}\bra{\psi_0}]$ is simply the overlap between $\ket{\psi(t)}$ and $\ket{\psi_0}$. When $\ket{\psi_0)}$ is an approximation of the ground state with a non-vanishing overlap, we can efficiently find the ground state energy, which corresponds to  the highest peak of $C(E,0)$. We note that the energy evaluation method correspond to the one introduced in Ref.~\cite{zeng2021universal}. 

\section{Degenerate energy levels}
When dealing with systems with degenerate energy levels, several peaks appear at the same place and make the peak higher. This in theory will not affect the result, instead it makes the peak better recognizable. In practice, however, these several peaks may not overlap perfectly due to errors and noise, resulting in several peaks which are very close to each other and thus making it hard to distinguish whether they refer to a degenerate energy or several different energy levels. Degeneracy stems from some symmetry of the system. Therefore, for a Hamiltonian with degenerate states, there must exist an operator $S$ satisfying $[S,H]=0$, thus the degenerate states are eigenstates of $S$ with different eigenvalues, i.e., $S\ket{\psi_l}=s_l\ket{\psi_l}$ where $s_l$ is different for each degenerate state $\ket{\psi_l}$. Therefore, one could further measure $p(t)\tr[OS\ket{\psi(t)}\bra{\psi(t)}]$ to identify degenerate energy levels with noisy experimental data when the shape of certain peaks has notably changed.

\section{Numerical simulation}

\subsection{Energy gap evaluation for the Heisenberg model}

To demonstrate the efficiency of our method, we consider a 4-qubit Heisenberg model and evaluate the energy gaps between the ground state and all other eigenstates. The Hamiltonian is given by
\begin{align}
H=-J\sum_{i=1}^{N_S-1}(X_iX_{i+1}+Y_iY_{i+1}+Z_iZ_{i+1})-h\sum_{i=1}^{N_S}Z_i,
\label{eq_Heisen}
\end{align}
 where the number of spins is $N_S = 4$, and $J=h=1$. $X_i$, $Y_i$, and $Z_i$ are the Pauli operators acting on the $i$-th site. We set the observable in Eq.~(\ref{eq:c(e)}) to be $O=\ket{\psi_0}\bra{\psi_0}$, where the state $\ket{\psi_0}$ represents the initial state. This leads to the transformation of Eq.~(\ref{eq:c(e)}) into
\begin{align}
C(E)&=\int_{-\infty}^\infty |\brakett{\psi_0}{\psi(t)}|^2e^{-a^2t^2}e^{iEt}dt.
\label{eq:c(e)trans}
\end{align} 
To evaluate $C(E)$ in Eq.~(\ref{eq:c(e)trans}), we employ the Monte Carlo method. As the Monte Carlo summation comes with a finite variance depending on the sampling approach, we apply the importance sampling method, which involves generating random samples of the evolution time $t$ that follow a Gaussian distribution function $p(t) \propto e^{-a^2t^2}$ on a classical computer. This helps to minimize the variance. The formula of the Monte Carlo summation is given by
\begin{align}
C(E)\approx\frac{1}{N_t}\sum_{s=1}^{N_t}\frac{f(t_s)}{p(t_s)},
\label{eq:sum}
\end{align}
where $N_t$ denotes the number of samples and $f(t_s)=|\brakett{\psi_0}{\psi(t_s)}|^2e^{-a^2t_s^2}e^{iEt_s}$. Subsequently, we obtain $|\brakett{\psi_0}{\psi(t)}|^2$ using a classical emulator of a quantum computer and finally evaluate the entire integral. 

\begin{figure}[t]
\centering\includegraphics[width=1\hsize]{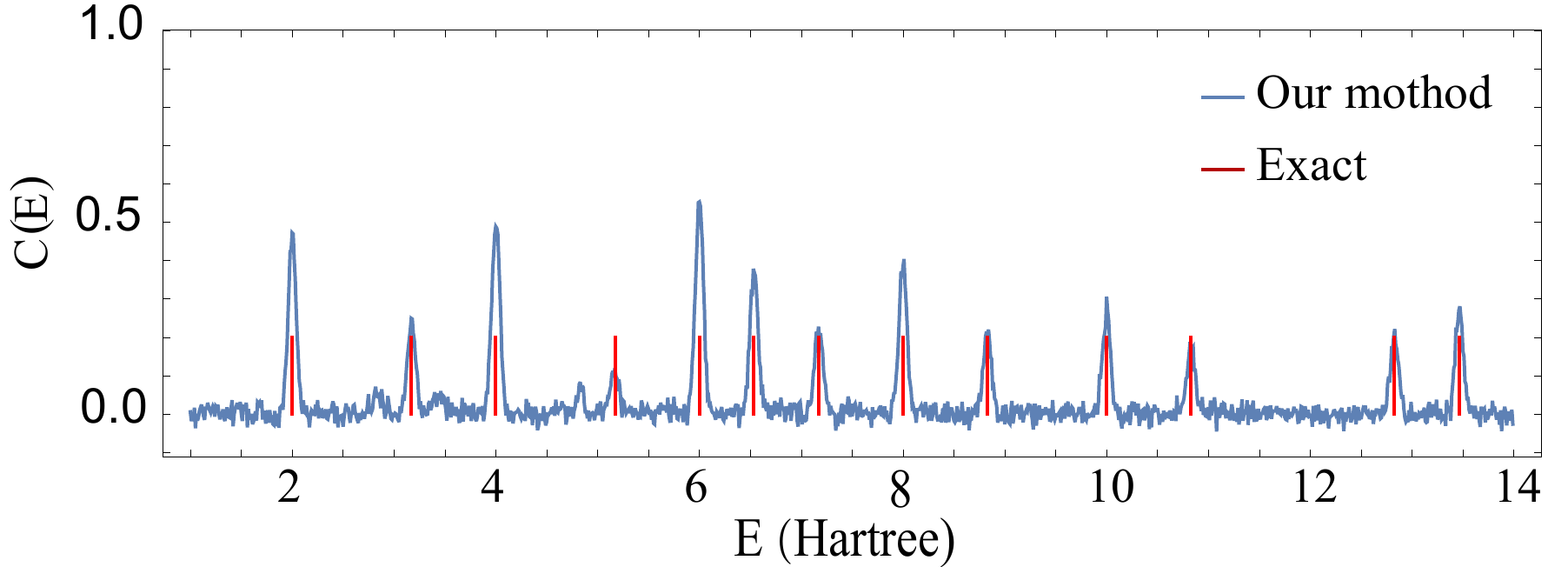}
\caption{Energy gaps of the 4-qubit Heisenberg model. The energy gap is defined as the energy difference between the ground state and other eigenstates. The red line in the figure indicates the exact value of the energy gap, while the light blue line represents $C(E)$ as obtained by our method.}
\label{fig:1}
\end{figure}

The results are shown in Fig.~\ref{fig:1}. We can see that $C(E)$ reaches a maximum value when $E=E_i-E_0$, where $E_0$ and $E_i$ denote the ground state energy and the energy of the $i$-th eigenstate, respectively. With degenerate eigenstates, in total 13 peaks are visible in the figure. The height of a peak depends on the overlap between the initial state and the corresponding eigenstate, as well as the level of degeneracy. We set the parameter $a=1/(25\sqrt{2})$ and sample 10000 times using the Monte Carlo method. The initial state is chosen as a superposition state of a randomly generated basis state, like $\ket{+++-}$, and the HF state. The process is repeated 10 times, and an averaged result is taken. As we can see from the figure, the location of the peaks matches well with the exact value of the energy gaps.

Note that the choice of the initial state $\ket{\psi_0}$ can significantly impact the results obtained through our method. For small quantum systems, selecting certain initial states that have a reasonable overlap with all eigenstates can lead to valid energy gaps between the ground state and other eigenstates. However, as the system size increases, the number of eigenstates of a quantum system grows exponentially, making it practically impossible to identify such states. Fortunately, in most cases, the energy gaps between the ground state and low-lying excited states are of primary interest. In such cases, we can prepare an initial state that has a reasonable overlap with the ground state and low-lying excited states. This enables us to evaluate the energy gaps between these states. Below, we present an example of such a case for the H$_4$ molecule.

\subsection{Energy gap evaluation for the H$_4$ molecule}

\begin{figure}[t]
\centering\includegraphics[width=1\hsize]{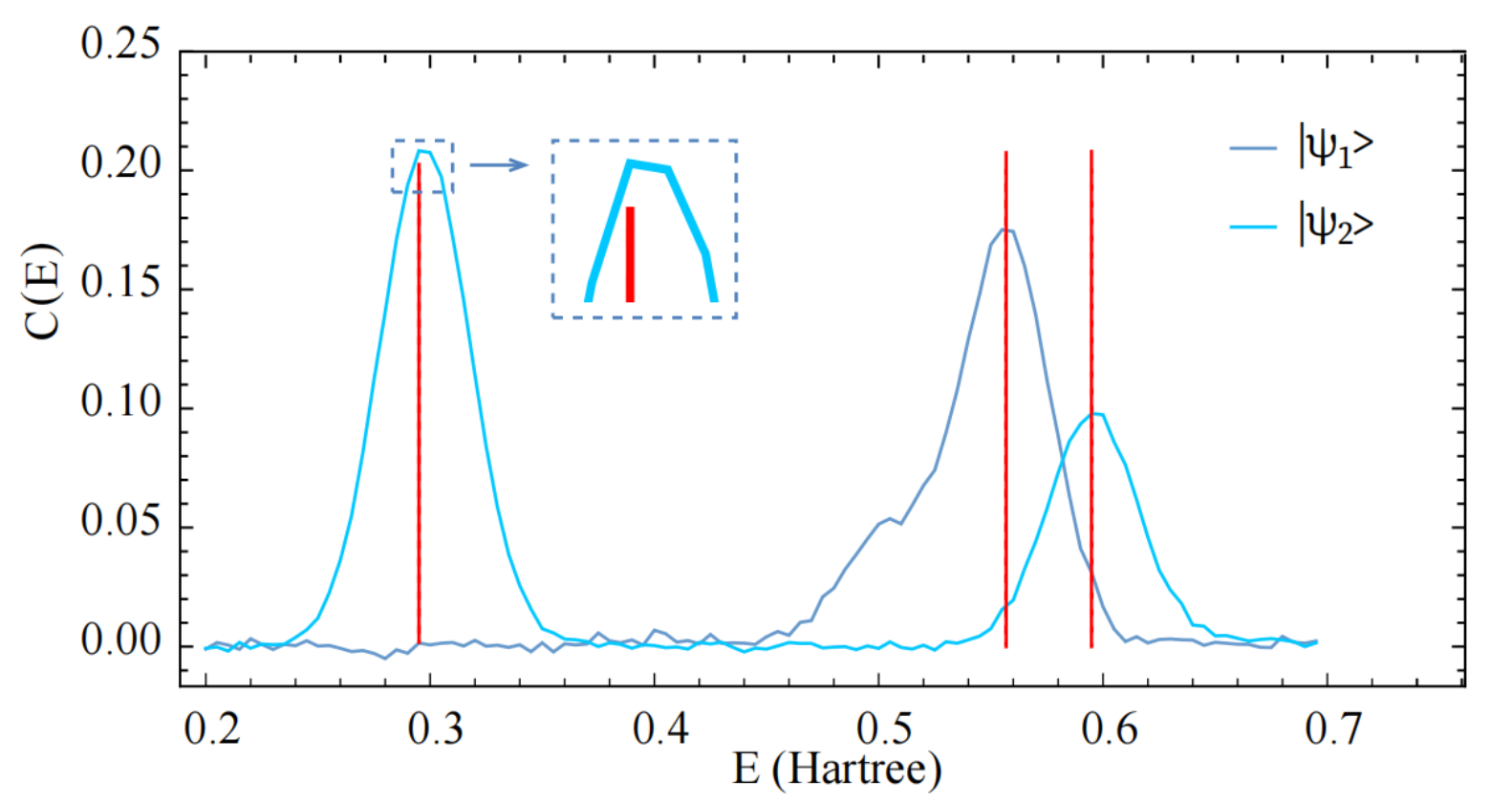}
\caption{The energy gaps between the ground state and low-lying energy states of the H$_4$ molecule. In the figure, the red line denotes the exact value of the energy gap, while the light blue and green lines represent the $C(E)$ values obtained using two initial states with our method.}
\label{fig:2}
\end{figure}

In this section, we consider a one-dimensional open-chain H$_4$ molecule with a bond length of 0.89 angstroms. The Hamiltonian is obtained in a second-quantized form using OpenFermion~\cite{mcclean2020openfermion}, an open-source quantum chemistry package. The Hamiltonian can be encoded using an 8-qubit quantum circuit. We then evaluate the energy gaps between the ground state and several low-lying excited eigenstates by calculating $C(E)$ using Eq.~(\ref{eq:c(e)trans}). The process is the same as described above, and the results are presented in Fig.~\ref{fig:2}.

To ensure that the initial state has a reasonable overlap with the low-lying excited eigenstates, we can generate the initial states by applying different single or double excitation operators on the Hartree–Fock state. In this example, we choose two states, $\ket{\psi_1}=U_1\ket{HF}$ and $\ket{\psi_2}=U_2\ket{HF}$, as the initial states. Here, $\ket{HF}$ denotes the Hartree–Fock state of the H$_4$ molecule, and $U_1=e^{a_7 a_4 ^ \dag - a_4 a_7 ^ \dag}$ and $U_2=e^{a_6 a_4 ^ \dag - a_4 a_6 ^ \dag}$ are two single excitation operators. Here, $a_i$ and $a_i ^ \dag$ are the annihilation operator and the creation operator acting on the $i$-th qubit, respectively. The choice of $U_1$ and $U_2$ is based on the mean-field approximation method used to solve the low-lying eigenstates of a quantum many-body system.

 In Fig.~\ref{fig:2}, we find that $C(E)$ reaches a peak when $E=E_i-E_0$, where $E_0$ is the ground state energy and $E_i$ is the energy of the $i$-th eigenstate. The height of a peak depends on the overlap between the initial state and the corresponding eigenstate. We set the parameter $a=1/(50\sqrt{2})$ and sample 50000 times using Monte Carlo method. As shown in Fig.~\ref{fig:2}, the location of peaks matches well with the the exact value of energy gaps.

\subsection{Energy evaluation}

\begin{figure}[t]
\centering\includegraphics[width=1\hsize]{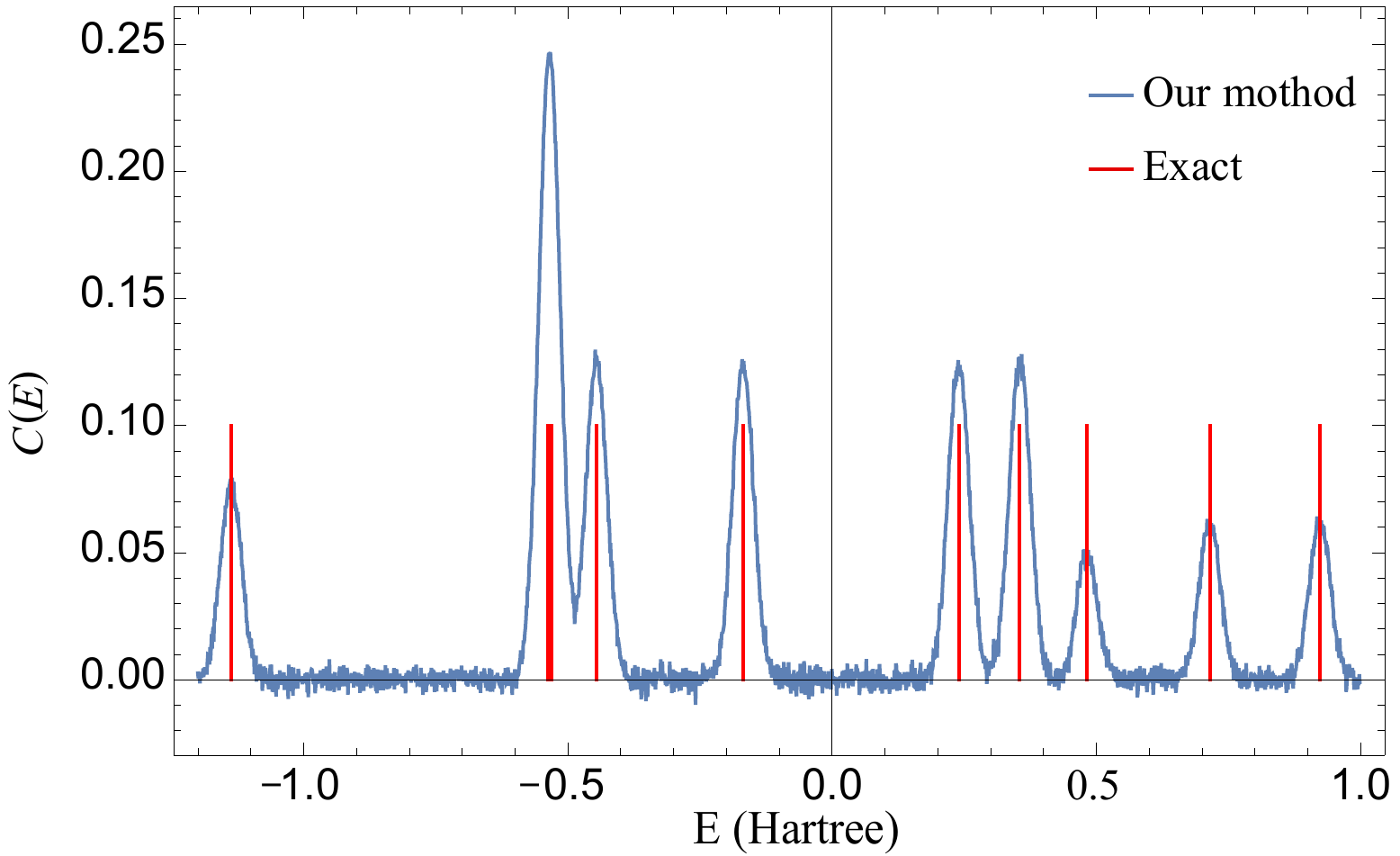}
\caption{The energy eigenvalues of a H$_2$ molecule, as determined by our method. In the figure, the red line indicates the exact value of the eigenstate energy, while the light blue line represents $C(E)$ as evaluated with our method.}
\label{fig:3}
\end{figure}

To evaluate the eigenenergies of a Hamiltonian, we can use Eq.~(\ref{eq:energy_direct}) with $O=I$. Then, the equation is transformed into 
\begin{align}
C(E)&=\int_{-\infty}^\infty\brakett{\psi_0}{\psi(t)}e^{-a^2t^2}e^{iEt}dt
\label{eq9}
\end{align}
We can use a quantum computer to measure $\brakett{\psi_0}{\psi(t)}$ in Eq.~(\ref{eq9}) and evaluate the integral with a classical computer using the Monte-Carlo method. We perform simulations for a range of molecules including H$_2$, H$_4$, LiH, CH$_2$ and NH. Particularly, we obtain the whole energy eigenvalues of a H$_2$ molecule in sto-3g basis, as shown in Fig.~\ref{fig:3}. For larger molecules, we evaluate their ground states, and the results are shown in Table~\ref{table1}.

In Fig.~\ref{fig:3}, we consider the H$_2$ molecule with a bond length of 0.74 angstrom. The initial state is choosen to be $\ket{---+}$. As shown in the figure, $C(E)$ reaches a peak when $E=E_i$, where $E_i$ denotes the $i$-th eigenstate energy. The parameter $a$ is fixed to be $1/(50\sqrt{2})$, and the number of samples is 10000.
In Table~\ref{table1}, $E_0$ is the exact ground-state energy of molecules, and $E_0^{'}$ is the energy obtained with our method. We find that our method can find the ground state of molecules with high accuracy.

\begin{table}[t]   
\begin{center}    
\begin{tabular}{|c|c|c|c|}   
\hline   Systems & $E_0^{'}$/Hartree & $E_0$/Hartree & Basis \\
\hline   H$_2$   & $-1.137$ & $-1.137284$  & sto-3g\\ 
\hline   H$_4$   & $-2.181$ & $-2.180501$  & sto-3g\\ 
\hline   LiH   & $-7.864$ & $-7.864266$  & sto-3g\\  
\hline   NH    &$-54.950$ & $-54.95045$   & 6-31g\\
\hline   CH$_2$  &$-38.904$& $-38.904354$  & 6-31g \\ 
\hline   
\end{tabular}   
\end{center}   
\caption{The ground state energy of multiple molecules. $E_0$ represents the exact ground-state energy, and $E_0^{'}$ denotes the energy obtained by our method.}
\label{table1}
\end{table}

\section{Error analysis}
Using our method, one applies Monte Carlo sampling to obtain  $C(E)=\int_{-\infty}^\infty p(t)\tr[O\ket{\psi(t)}\bra{\psi(t)}]e^{iEt}dt$  and effectively evaluates
\begin{align}
C(E)=\sum_{n=-N}^N p(n\tau) \tr[O\ket{\psi(n\tau)}\bra{\psi(n\tau)}]e^{iEn\tau},
\label{Eq.c(e)_discrete}
\end{align}
and the averaged results generate a curve with peaks appearing at $E=E_i-E_j=\Delta_{ij}$. Whether one can accurately determine $\Delta_{ij}$ depends on two factors: (1) The position of the peak is accurate; (2) The width of the peaks is small enough that one can distinguish two neighboring peaks. 
For the first factor, the accuracy of the location of a certain peak is mostly affected by the peaks next to it. For example, when evaluating the ground state, the accuracy scales with the energy gap of the ground and first excited states (details can be found in Appendix~\ref{Accuracy}), thus our method is most suitable for systems with finite energy levels.

\begin{figure}[t]
\centering\includegraphics[width=1\hsize]{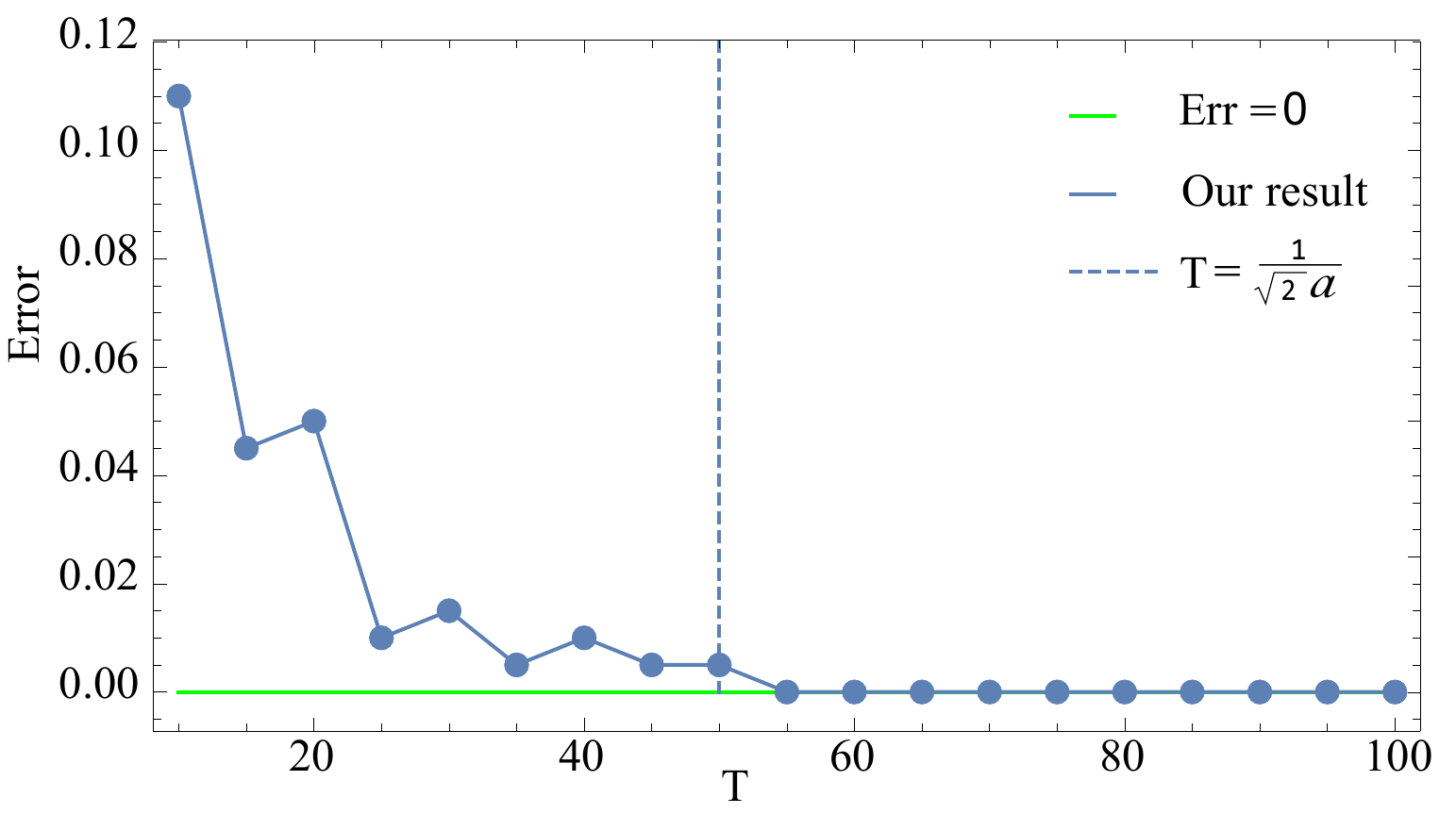}
\caption{The error in the energy gap between the ground state and the first excited state of the H$_2$ molecule, as a function of the sampling range. The light blue line represents the error caused by the cutoff, while the green line indicates the error-free case. The position of the dashed line marks the lower bound of the cutoff time.}
\label{fig:4}
\end{figure}

The shape of the curve depends on the Fourier transform of function $p(t)$, and in our case, it is a Gaussian function $e^{-\frac{(E-E_i)^2}{4a^2}}$. In the frequency domain, we want to make the curve around each peak sharp to better locate the crest, i.e., a small $a$ is preferred, this on the other hand leads to a broader curve distribution in the time domain. To limit the evolution time, we can use a cutoff of the sampling range from $[-\infty,\infty]$ to $[-T,T]$ when evaluating Eq.~(\ref{Eq.c(e)_discrete}). We define $C(E)^\infty$ and $C(E)^T$ as obtained from an infinite and a cutoff range, respectively. The cutoff error $\epsilon=|C(E)^\infty - C(E)^T|$ can be shown to be $\epsilon\leqslant\frac{2}{a}e^{-a^2T^2}$. Therefore, if we want to constrain the error $\epsilon$ to $\epsilon_c$, the sampling range should be no smaller than $\frac{1}{a}\sqrt{\rm{In\frac{2}{a\epsilon_c}}}$. (Details are provided in Appendix~\ref{error_analysis}).

In Fig.~\ref{fig:4}, we illustrate how the error changes with the sampling range $T$. Here, the error is defined as the difference between the exact energy gap value and the value with a cutoff when evaluating the energy gap between the ground state and the first excited state of the H$_2$ molecule. We fix the parameter $a=1/(50\sqrt{2})$, and the number of sampling times using the Monte Carlo method is 50000. As shown in Fig.~\ref{fig:4}, when the sampling range $T$ is greater than 50, the error decreases rapidly to zero.

When measuring $\tr[O\ket{\psi(n\tau)}\bra{\psi(n\tau)}]$ with a quantum computer, one can decompose $O$ into Pauli operators $O=\sum_i\alpha_iP_i$ and measure each term. The sampling noise exists with a variance bounded by $\frac{1}{4N_s}\sum_i|\alpha_i|^2$ (details given in Appendix~\ref{error_analysis}). 

\section{Discussion and Conclusion}

In this work, we introduce a new hybrid non-variational algorithm to evaluate energy gaps for a given Hamiltonian $H$. In our algorithm, we use real-time evolution simulation to obtain the value of function $f(t)$ on a quantum computer and evaluate the entire Fourier integral with a classical computer. The Fourier transform is defined by $C(E)=\int_{-\infty}^\infty f(t)e^{iEt}dt$, where the specific form of function $f(t)$ depends on the observable that we want to evaluate. Here, when $E$ equals a certain energy gap or an eigenvalue, depending on the problem, $C(E)$ reaches a maximum.

During the process, as the entire evolution simulation is performed on a quantum computer, our algorithm is free from the sign problem. Furthermore, the evolution time is constrained as we consider a cutoff of the sampling range from $[-\infty,\infty]$ to $[-T,T]$ to reduce the circuit depth. And we show the energy gap error can be close to zero when a reasonable $T$ is taken. Therefore, our method can be implemented with shallow circuits and is suitable for near-term devices.

\textbf{Note added.} 
During the preparation of this work, we became aware 
of two relevant works, one posted some months ago~\cite{chan2023algorithmic} and another very recent one~\cite{jinzhao} that consider similar problems and similar approaches for finding energy gaps using Fourier transform of results from real time evolution. 

\bibliography{reference}

\appendix

\section{\label{Accuracy}Accuracy of the location of the peaks}
In this section we discuss the accuracy of the peaks appearing at the correct location. We first consider the case to evaluate the energy using Eq.~(\ref{eq:energy_direct}).
The peaks in the curve of $C(E)$ occur at $\frac{\partial C(E)}{\partial E}=0$:
\small
\begin{align}
&\frac{\partial C(E)}{\partial E}=\frac{\partial}{\partial  E}\left(\tr[O\sum_{i=0}\ket{i}\brakett{i}{\psi_0}\bra{\psi_0}\frac{\sqrt{\pi}}{a}e^{-\frac{(E-E_i)^2}{4a^2}}]\right) \nonumber \\
=&-\tr[O\sum_{i=0}\ket{i}\brakett{i}{\psi_0}\bra{\psi_0}\frac{\sqrt{\pi}(E-E_i)}{2a^3}e^{-\frac{(E-E_i)^2}{4a^2}}]=0.
\end{align}
\normalsize
Let's write $\ket{i}\brakett{i}{\psi_0}\bra{\psi_0}=D_i$, so we have
\begin{align}
\sum_{i=0}D_i(E-E_i)e^{-\frac{(E-E_i)^2}{4a^2}}=0.
\label{eq:ap:single_energy_peak}
\end{align}
In most applications we want to find the ground state $E_0$, so we consider the extreme point around $E=E_0$. As the contribution from peaks far away from $E_0$ scales down exponentially, the peak of $E_1$ attributes the largest contribution to Eq.~(\ref{eq:ap:single_energy_peak}):

\begin{align}
&D_0(E-E_0)e^{-\frac{(E-E_0)^2}{4a^2}}+D_1(E-E_1)e^{-\frac{(E-E_1)^2}{4a^2}}=0. \nonumber \\
&\Longrightarrow E=\frac{D_0E_0e^{-\frac{(E-E_0)^2}{4a^2}}+D_1E_1e^{-\frac{(E-E_1)^2}{4a^2}}}{D_0e^{-\frac{(E-E_0)^2}{4a^2}}+D_1e^{-\frac{(E-E_1)^2}{4a^2}}}
\end{align}
Therefore, the error of the location is bounded by:
 \begin{align}
|E-E_0|\approx\left|\frac{D_1(E_1-E_0)e^{-\frac{(E_0-E_1)^2}{4a^2}}}{D_0+D_1e^{-\frac{(E_0-E_1)^2}{4a^2}}}\right|<\frac{D_1(E_1-E_0)}{D_0},
\end{align}   
as $D_0>D_1$. We can see that the accuracy of the result depends on the gap of the ground state and the first excite state. Therefore, for physical systems with finite energy gaps,e.g., molecules, our method can be quite accurate.

Next, we consider the case when evaluating the energy gaps. The peaks in the curve of $C(E)$ occur at $\frac{\partial C(E)}{\partial E}=0$:
\small
\begin{widetext}
\begin{align}
\frac{\partial C(E)}{\partial E}=&\frac{\partial}{\partial E}(\tr[O\sum_{i,j=0}\ket{i}\brakett{i}{\psi_0}\brakett{\psi_0}{j}\bra{j}\frac{\sqrt{\pi}}{a}e^{-\frac{(E-E_i-E_j)^2}{4a^2}}]) \nonumber \\
=&\tr[O\sum_{i,j=0}\ket{i}\brakett{i}{\psi_0}\brakett{\psi_0}{j}\bra{j}-\frac{\sqrt{\pi}(E-E_i-E_j)}{2a^3}e^{-\frac{(E-E_i-E_j)^2}{4a^2}}].
\end{align}
\end{widetext}
\normalsize
Let's write $\ket{i}\brakett{i}{\psi_0}\brakett{\psi_0}{j}\bra{j}=D_{i,j}$ and $E_i-E_j=\Delta_{i,j}$, then the above function changes into:
\begin{align}
\frac{\partial C(E)}{\partial E}=-\tr[O\sum_{i,j=0}D_{i,j}\frac{\sqrt{\pi}(E-\Delta_{i,j})}{2a^3}e^{-\frac{(E-\Delta_{i,j})^2}{4a^2}}]=0,
\end{align}
then
\begin{align}
\sum_{i,j=0}D_{i,j}(E-\Delta_{i,j})e^{-\frac{(E-\Delta_{i,j})^2}{4a^2}}=0.
\end{align}
We focus on the peak around $E=\Delta_{i,j}$ and consider a general case that the peak has two neighboring peaks at $E=\Delta_1$ and $E=\Delta_2$, with $\Delta_1<\Delta_{i,j}<\Delta_1$. Therefore, the location of the target peak is mostly affected by these two peaks:
\small
\begin{align}
&D_{i,j}(E-\Delta_{i,j})e^{-\frac{(E-\Delta_{i,j})^2}{4a^2}}+D_1(E-\Delta_1)e^{-\frac{(E-\Delta_1)^2}{4a^2}}+\notag\\
&D_2(E-\Delta_2)e^{-\frac{(E-\Delta_2)^2}{4a^2}}=0 \nonumber \\
&\Longrightarrow\notag\\
&E=\frac{D_1\Delta_1e^{-\frac{(E-\Delta_1)^2}{4a^2}}+D_2\Delta_2e^{-\frac{(E-\Delta_2)^2}{4a^2}}+D_{i,j}\Delta_{i,j}e^{-\frac{(E-\Delta_{i,j})^2}{4a^2}}}{D_1e^{-\frac{(E-\Delta_1)^2}{4a^2}}+D_2e^{-\frac{(E-\Delta_2)^2}{4a^2}}+D_{i,j}e^{-\frac{(E-\Delta_{i,j})^2}{4a^2}}}.
\end{align}
\normalsize
Therefore, the accuracy is bounded by
\begin{align}
\small
&|E-\Delta_{i,j}|\approx\notag\\
&\left|\frac{-D_1(\Delta_{i,j}-\Delta_1)e^{-\frac{(\Delta_{i,j}-\Delta_1)^2}{4a^2}}+D_2(\Delta_2-\Delta_{i,j})e^{-\frac{(\Delta_{i,j}-\Delta_2)^2}{4a^2}}}{D_1e^{-\frac{(\Delta_{i,j}-\Delta_1)^2}{4a^2}}+D_2e^{-\frac{(\Delta_{i,j}-\Delta_2)^2}{4a^2}}+D_{i,j}}\right|.
\normalsize
\end{align}
$|E-\Delta_{i,j}|$ is mostly affected by the nearest neighboring peak. Besides, in any case, we have $|E-\Delta_{i,j}|<\Delta_{i,j}-\Delta_1$ and $|E-\Delta_{i,j}|<\Delta_2-\Delta_{i,j}$ so that all the peaks can be distinguished from each other.

\section{\label{error_analysis}Error analysis}
We consider two dominant error sources that affect the precision of our method: cutoff error and sampling error. 
\subsection{Cutoff error}
Using our method, $C(E)$ is evaluated using the Monte Carlo method. In practice, one cannot sample from minus infinity to infinity. Here we consider a cutoff to the sampling range from $[-\infty,\infty]$ to $[-T,T]$. Then, we can define
\begin{align}
\small
&C(E)^\infty=\notag\\
&\tr[O\sum_{i,j=0}\ket{i}\brakett{i}{\psi_0}\brakett{\psi_0}{j}\bra{j}\int_{-\infty}^\infty e^{-i(E_i-E_j-E)t}*p(t)dt] \nonumber \\
&=\tr[O\sum_{i,j=0}\ket{i}\brakett{i}{\psi_0}\brakett{\psi_0}{j}\bra{j}G(E_{i,j})],
\normalsize
\end{align}
where we make $G(E_{i,j})^\infty=\int_{-\infty}^\infty e^{-i(E_i-E_j-E)t}*p(t)dt$, $p(t)=e^{-a^2t^2}$ and $O=\ket{\psi_0}\bra{\psi_0}$. Then the cutoff error is
\begin{align}
&|C(E_{i,j})^\infty - C(E_{i,j})^T|\leq|G(E_{i,j})^\infty - G(E_{i,j})^T| \nonumber \\
=&\int_{-\infty}^\infty e^{-i(\Delta_{i,j}-E)t}p(t)dt -\int_{-T}^T e^{-i(\Delta_{i,j}-E)t}p(t)dt \nonumber \\
=&\int_{-\infty}^{-T} e^{-i(\Delta_{i,j}-E)t}p(t)dt+\int_{T}^\infty e^{-i(\Delta_{i,j}-E)t}p(t)dt \nonumber \\
\leqslant&\int_{-\infty}^{-T} p(t)dt + \int_{T}^\infty p(t)dt \nonumber \\
=&\frac{2}{a}\rm{erfc}(aT).
\end{align}
$\rm{erfc}(x)$ is the error function which satisfies $\rm{erfc}(x)\leqslant e^{-x^2}$. For $x\geqslant0$, $|C(E_{i,j})^\infty - C(E_{i,j})^T|\leqslant\frac{2}{a}e^{-a^2T^2}$. Therefore, if we want to constrain the error by $\epsilon_c$, the sample range should be no smaller than $\frac{1}{a}\sqrt{\rm{In\frac{2}{a\epsilon_c}}}$.

\subsection{Sampling noise}
As described in the main text, we take finite samples to generate $C(E)$ using the Monte Carlo method:
\begin{align}
C(E)=\sum_{n=-N}^N \tr[O\ket{\psi(n\tau)}\bra{\psi(n\tau)}*p(n\tau)]e^{iEn\tau}\tau,
\end{align}
In each sampling, one need to measure $\tr[O\ket{\psi(n\tau)}\bra{\psi(n\tau)}]$ using a quantum computer. To do so, we can decompose $O$ into Pauli operators $O=\sum_i \alpha_iP_i$ where $P_i$ is a Pauli operator. Then one prepares $\ket{\psi(n\tau)}$ and measures $\langle P_i\rangle$ each time. In each shot, one gets a binary number of $u=\pm 1$ as the result. Therefore, the variance of $\langle P_i\rangle$ is given by 
\begin{align}
\rm{Var}[P_i]=\frac{1-\mathbf{E}[u]^2}{4N_s}\leqslant\frac{1}{4N_s},
\end{align}
where $N_s$ is the number of samples. So the variance of $\tr[O\ket{\psi(n\tau)}\bra{\psi(n\tau)}]$ is bounded by $\frac{1}{4N_s}\sum_i|\alpha_i|^2$.

\end{document}